# Machine-learning-based Classification of Lower-grade gliomas and High-grade gliomas using Radiomic Features in Multi-parametric MRI

Ge Cui[1]; Jiwoong Jeong[1]; Bob Press[1]; Yang Lei[1];

Hui-Kuo Shu[1]; Tian Liu[1]; Walter Curran[1]; Hui Mao[1,2]; Xiaofeng Yang[1*]

[1]Department of Radiation Oncology and Winship Cancer Institute, Emory University, Atlanta, GA 30322

[2]Department of Radiology and Imaging Sciences and Winship Cancer Institute, Emory University, Atlanta, GA 30322

*Corresponding to xiaofeng.yang@emory.edu

**Abstract**

**Objectives:** Glioblastomas are the most aggressive brain and central nervous system (CNS) tumors with poor prognosis in adults. The purpose of this study is to develop a machine-learning based classification method using radiomic features of multi-parametric MRI to classify high-grade gliomas (HGG) and low-grade gliomas (LGG).

**Methods:** Multi-parametric MRI of 80 patients, 40 HGG and 40 LGG, with gliomas from the MICCAI BRATs 2015 training database were used in this study. Each patient's T1, contrast-enhanced T1, T2, and Fluid Attenuated Inversion Recovery (FLAIR) MRIs as well as the tumor contours were provided in the database. Using the given contours, radiomic features from all four multi-parametric MRIs were extracted. Of these features, a feature selection process using two-sample T-test and least absolute shrinkage, selection operator (LASSO), and a feature correlation threshold was applied to various combinations of T1, contrast-enhanced T1, T2, and FLAIR MRIs separately. These selected features were then used to train, test, and cross-validate a random forest to differentiate HGG and LGG. Finally, the classification accuracy and area under the curve (AUC) were used to evaluate the classification method.

**Results:** Optimized parameters showed that on average, the overall accuracy of our classification method was 0.913 or 73 out of 80 correct classifications, 36/40 for HGG and 37/40 for LGG, with an AUC of 0.956 based on the combination with FLAIR, T1, T1c and T2 MRIs.

**Conclusion:** This study shows that radiomic features derived from multi-parametric MRI could be used to accurately classify high and lower grade gliomas. The radiomic features from multi-parametric MRI in combination with even more advanced machine learning methods may further elucidate the underlying tumor biology and response to therapy.

# 1. INTRODUCTION

Gliomas are the most common, aggressive, and lethal primary brain tumors, accounting for nearly 33% of all brain tumors [1]. Lower grade gliomas, grades 1-3, have an average relative 2 year survival rate of 80% while the relative 2 year survival for high grade gliomas, grade 4, drops to 30% with a median overall survival rate of 12 to 15 months [2, 3]. These poor survival rates stem from the heterogeneous nature of gliomas [4, 5]. Recently studies have demonstrated the predictive power of genetic characterization and expression of gliomas with respect to treatment response [5-10]. However, genetic and molecular heterogeneity of gliomas still pose an obstruction to developing novel treatments and clinical decision making [11]. This shows that there is a need for an accurate characterization and classification of these high and low grade gliomas for the selection of proper clinical decision making to improve the outcomes of these patients.

In recent years, non-invasive methods of characterizing tumors through a radiomic approach [12, 13], to avoid the complexities of intra-tumoral and inter-patient issues via sample-limited biopsies [14], has been growing. These radiomic approaches have the advantage of being able to interrogate the whole tumor without being invasive and has been correlated to a wide variety of clinical outcomes in the breast [15], brain [16], and head and neck [17] using different modalities like computer tomography (CT), positron emission tomography (PET), and magnetic resonance imaging (MRI).

Previous studies using single-sequence imaging were reasonably useful for glioma grading [18-20]. However, features derived from a single modality cannot comprehensively depict the heterogeneity of gliomas [21], which is an important characteristic for prognostic predictions [19].Thus, brain tumor patients with gliomas typically undergo multiple MRI scans to classify and determine the tumor type and grade as well as monitoring the treatment response and tumor recurrence. As mentioned before, histopathologic analysis of tumor tissue samples collected from either biopsy or surgical resection does not fully describe intra-tumoral heterogeneity due to its difficulty to repeatedly measure and limitations on sampling. In contrast, MRI exams with different imaging techniques, or parameters, could be performed non-invasively and frequently; effectively reducing patient burden and giving the ability to gather data during disease progression. In addition, MRI exams provide morphological and physiological information of the tumor with great spatial resolution and tissue characteristics. The four most common MRI exams given to patients are T1 weighted, T2 weighted, contrast enhanced T1 (T1c), and fluid attenuated inversion recovery (FLAIR) MRIs for their ability to create contrast based on MR relaxation rates to maximize anatomical or pathological imaging.

The purpose of this work is to improve the tissue classification of these highly heterogeneous gliomas by taking advantage of: 1) the different contrast enhancements of multi-parametric MRIs; 2) the radiomic features derived from each MRI; and 3) the random forest machine learning method that splits observations with similar response variables.

## 2. MATERIALS AND METHOD

The proposed classification method used in this study is described as in Figure 1. First, multi-parametric MR images are acquired, registered, and contoured. Then radiomic features are extracted. Once radiomic features were extracted, a feature selection method is applied separately to get the most pertinent features for various combinations of T1, T1c, T2, and FLAIR MRIs separately. Then selected features from various modality combinations were used to train, test, and cross-validate a random forest (RF) decision tree [22]. Finally, the performance of our classification method was evaluated and optimized accordingly.

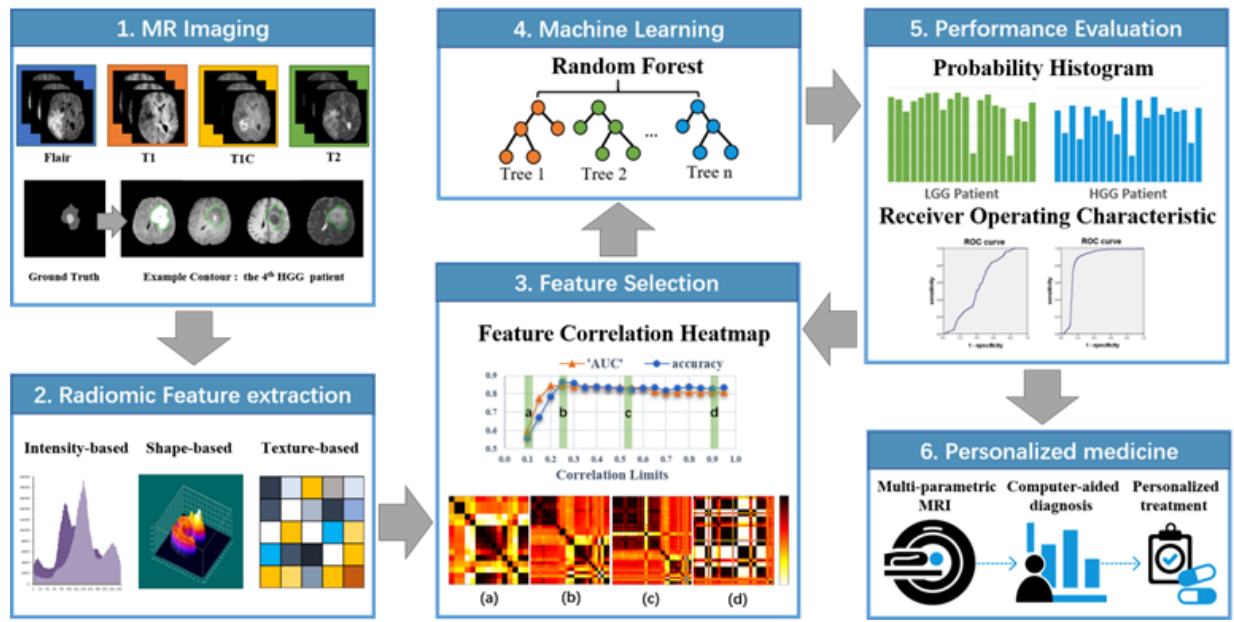

**Figure 1.** The Flowchart of our proposed classification method. This flowchart outlines the various steps in our classification method starting from image acquisition to performance evaluation.

### 2.1 Image Acquisition

40 high and 40 low grade patient image data for this study was randomly selected from the MICCAI BRATs 2015 training dataset. Low grade patients were defined as patients whose diseases were histologically diagnosed as astrocytomas or oligoastrocytomas while high grade patients were anaplastic astrocytomas and glioblastoma multiforme tumors. The multi-parametric MRIs were: 1) T1: T1-weighted, native image, sagittal or axial 2D acquisitions, with 1mm slice thickness; 2) T1c: T1-weighted, Gadolinium contrast-enhanced image, with 3D acquisition and 1 mm isotropic voxel size for most patients; 3) T2: T2-weighted image, axial 2D acquisition, with 1 mm slice thickness; 4) FLAIR: T2-weighted FLAIR image, axial, coronal, or sagittal 2D acquisitions, 1 mm slice thickness. The multi-parametric MR images were aligned to the same common space and interpolated to the 1x1x1mm voxel resolution with an image size of 240x240x155 [23].

## 2.2 Radiomic Feature Extraction

Using the multi-parametric MRIs and contours provided from the BRATs dataset, a total of 1730 features were extracted for each MR parameter (T1, T1c, T2, and FLAIR) for a total of 6920 features for each patient. The feature categories and their corresponding features referenced are found in the paper by Zhang *et al.* [24].

## 2.3 Feature Selection

Once all features were extracted for each patient, feature selection was applied. It is a procedure commonly applied in radiomic and machine learning studies that aim to extract the most informative features and reduce the redundant variables for computational efficiency. Our feature selection method involved three components: two tests of significance and one test of redundancy thresholds. The first test of significance, a two sample T-test, was to find those features that were significantly different between high and low grade patients. The second test, LASSO (least absolute shrinkage and selection operator), is a commonly used statistical regression analysis tool that is used to increase the prediction accuracy of statistical models by selecting unique and significant predictors and reduce variance without a substantial increase of a bias [25]. In this case, HGG and LGG patients were considered as two sample populations to select features that are significantly different at the 0.05 significance level and then the first round of LASSO was used to further select and reduce features. Only those features that were selected as significant by both tests, two sample T-test and LASSO, were chosen to be used for the next step. Once the significant features were selected in general, redundant features from various combinations of T1, T1c, T2, and FLAIR MRIs were removed separately using a correlation matrix-based selection with variable correlation thresholds. Figure 2 shows the heatmaps of the correlation matrix used to remove redundancies based on the combination of all the four MR modalities.

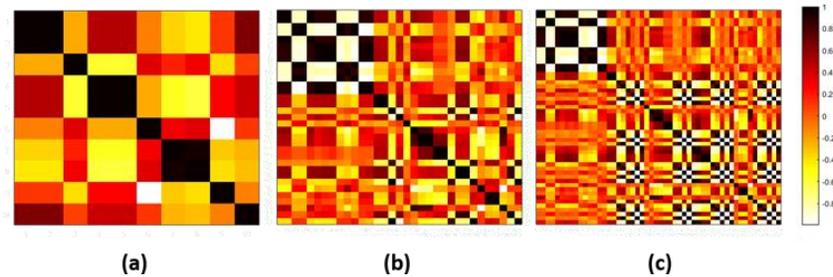

**Figure 2.** Heatmaps of feature correlation matrices at various correlation thresholds. The figure shows the threshold matrices, from left to right, less than (a) 0.2, (b) less than 0.3, and (c) less than 0.6 correlation.

**2.4 Machine Learning**

After feature selection, a RF machine learning method was used to train, test, and cross-validate a model for the classification of patients as HGG or LGG. While there are many well-known machine learning methods, this study chose RF for its advantages in accuracy, handling of large databases, generation of error, estimation of variable importance, and robustness to noise [26, 27]. Specifically, we chose the Matlab function, fit ensemble, with a semi-optimized template tree. At each correlation limit, the corresponding features and labels of each patient were supplied to the random forest to train and test it, producing various performance metrics such as prediction accuracy, prediction probability, and receiver operating characteristic (ROC) curves.

**2.5 Performance Evaluation**

In order to evaluate the performance of our classification method to the general population and reduce the overfitting bias, a leave-one-out cross-validation was applied to the random forest. This allowed our method to be trained and tested over many iterations, each with a new test patient. After the leave-one-out cross-validation testing, we evaluated the accuracy of our predictions through our criteria and calculating the area under the curve (AUC) of the ROC curve. In our approach, we also evaluated the performance of using one, two, three, and all four MR modalities.

### 3. RESULTS

The result for the variation of correlation thresholds from 0.10 to 0.95 at 0.025 increments using all four MR modalities is shown in Figure 3. The best overall accuracy, 0.913, and AUC, 0.956, were observed at the correlation threshold of 0.575. The best results achieved at various combinations of correlation thresholds and MRI modalities are shown in Table 1. At the correlation threshold of 0.575, the accuracy probability histogram of HGG and LGG patients are shown in Figure 4. Finally, a comparison of ROC curves between the worst and best prediction thresholds are shown in Figure 5. The AUCs are 0.757 and 0.956 at the correlation thresholds of 0.100 and 0.575 respectively.

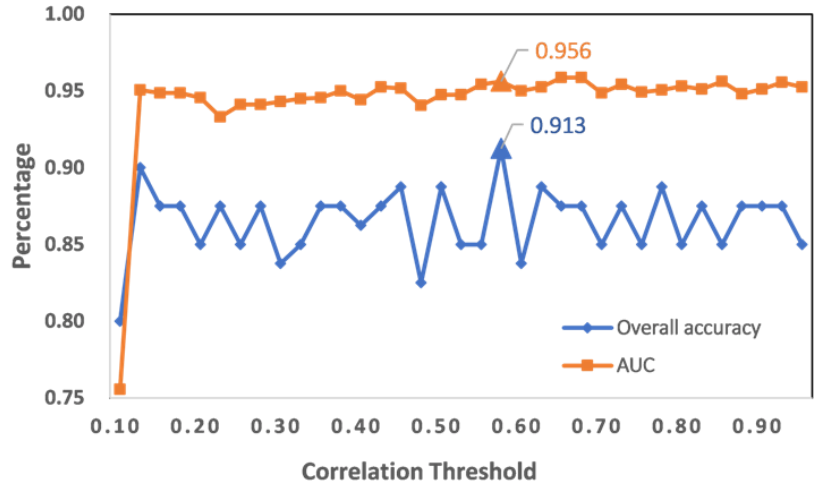

**Figure 3.** Overall accuracy and AUC over different correlation thresholds for using all modalities. This figure shows how the overall prediction accuracy and area under the curve (AUC) changes over different correlation thresholds for the best performing combination of modalities (all four).

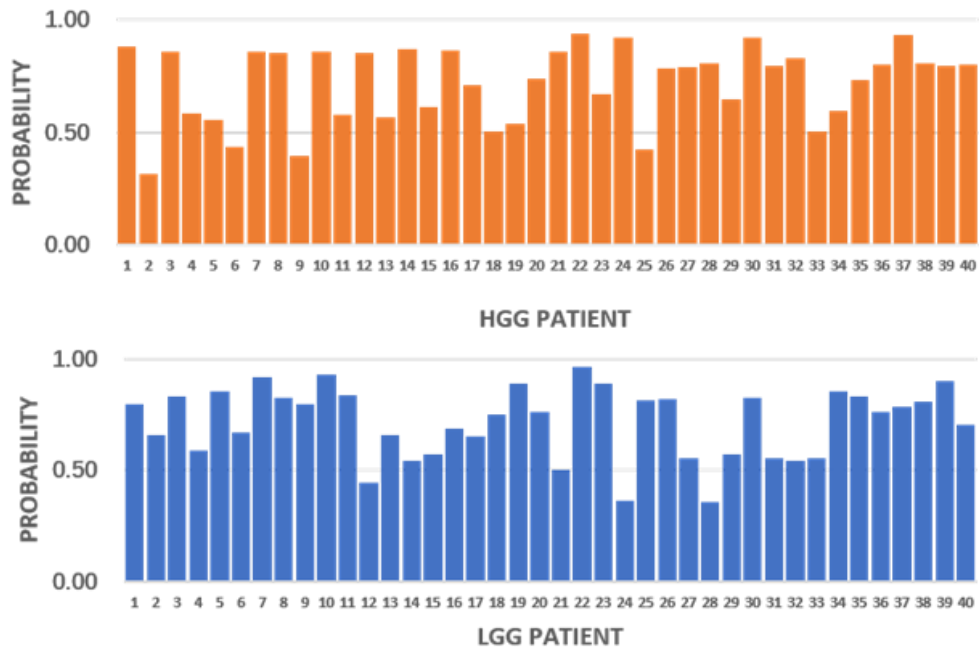

**Figure 4.** Predicted accuracy of each HGG and LGG patient using the leave-one-out cross-validation. This figure shows each patient's, high grade glioma (HGG) or low grade glioma (LGG), prediction accuracy at the best performing combination of modalities and threshold.

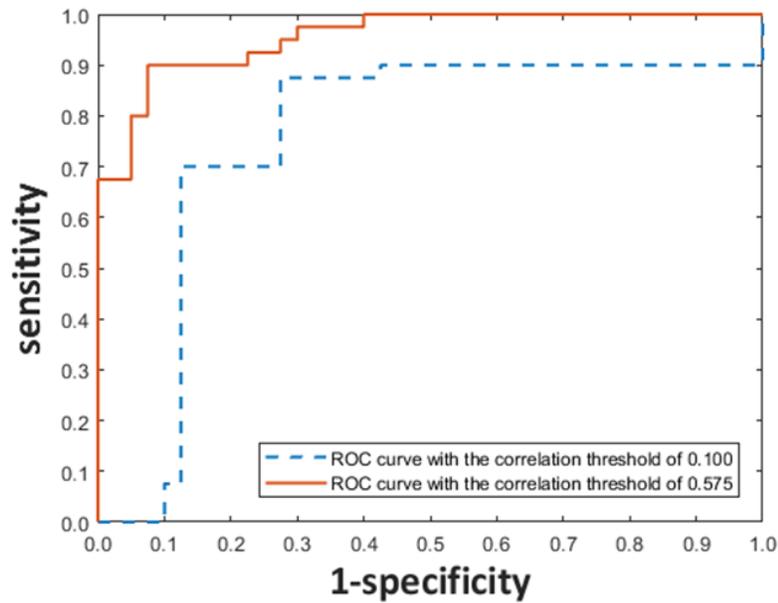

**Figure 5.** The AUC of the best correlation threshold (0.575) and the worst correlation threshold (0.100). A comparison of the area under the curve (AUC) of the best and worst correlation thresholds that outlines the need to fine tune feature selections.

**Table 1.** The best results for overall patients' accuracies and AUCs at various combination of modalities.

| Combination | Overall Accuracy | AUC |
| --- | --- | --- |
| Flair | 0.763 | 0.782 |
| T1 | 0.775 | 0.826 |
| T1c | 0.863 | 0.918 |
| T2 | 0.813 | 0.818 |
| Flair+T1 | 0.813 | 0.809 |
| Flair+T1c | 0.838 | 0.904 |
| Flair+T2 | 0.800 | 0.808 |
| T1+T1c | 0.850 | 0.899 |
| T1+T2 | 0.800 | 0.848 |
| T1c+T2 | 0.838 | 0.918 |
| T1+T1c+T2 | 0.875 | 0.907 |
| Flair+T1c+T2 | 0.838 | 0.870 |
| Flair+T1+T2 | 0.838 | 0.835 |
| Flair+T1+T1c | 0.863 | 0.913 |
| **Flair+T1+T1c+T2** | **0.913** | **0.956** |

**Table 2.** Features and feature groups for the best results at various combinations of modalities.

| Modality | Combination | Amount | Feature Group |
|---|---|---|---|
| Single modality | Flair | 36 | GLCM(22) Intensity(5) Shape(9) |
| | T1 | 15 | Intensity(3) Shape(12) |
| | T1c | 14 | Intensity(4) Shape(10) |
| | T2 | 11 | GLCM(2) Intensity(4) Shape(5) |
| Two modalities combined | Flair | 40 | GLCM(29) Intensity(5) Shape(6) |
| | T1 | 10 | GradientOrientHistogram(1) Intensity(3) Shape(6) |
| | Flair | 2 | Shape(2) |
| | T1c | 2 | Intensity(2) Shape(2) |
| | Flair | 11 | GLCM(7) Shape(4) |
| | T2 | 5 | GLCM(2) Shape(4) |
| | T1 | 3 | Shape(3) |
| | T1c | 5 | Intensity(2) Shape(3) |
| | T1 | 26 | GradientOrientHistogram(1) GLCM(11) GLRM(2) Intensity(5) Shape(7) |
| | T2 | 19 | GLCM(4) Intensity(7) Shape(8) |
| | T1c | 8 | Intensity(4) Shape(4) |
| | T2 | 8 | GLCM(3) Intensity(1) Shape(4) |
| Three modalities combined | Flair | 43 | GLCM(23) Intensity(5) Shape(5) |
| | T1 | 16 | GradientOrientHistogram(1) GLCM(7) Intensity(4) Shape(4) |
| | T1c | 8 | Intensity(4) Shape(4) |
| | Flair | 23 | GLCM(15) Intensity(3) Shape(5) |
| | T1c | 10 | Intensity(4) Shape(6) |
| | T2 | 10 | GLCM(3) Intensity(1) Shape(6) |
| | Flair | 29 | GLCM(20) Intensity(4) Shape(5) |
| | T1 | 9 | GradientOrientHistogram(1) Intensity(3) Shape(5) |
| | T2 | 10 | GLCM(2) Intensity(3) Shape(5) |
| | T1 | 12 | GLCM(7) Intensity(3) Shape(2) |
| | T1c | 5 | Intensity(3) Shape(2) |
| | T2 | 13 | GLCM(4) Intensity(6) Shape(3) |
| All modalities combined | Flair | 23 | GLCM(15) Intensity(3) Shape(5) |
| | T1 | 11 | GradientOrientHistogram(1) GLCM(3) Intensity(3) Shape(4) |
| | T1c | 8 | Intensity(4) Shape(4) |
| | T2 | 10 | GLCM(4) Intensity(2) Shape(4) |

# 4. DISCUSSION

## 4.1 The effects of multi-parametric predictions

While introducing more sources of information like multi-parametric MRI seems advantageous, there can be associated disadvantages as well. For example, adding more data increases computational burden, introduces known and unknown sources of error, and increases patient burden both financially and timewise. If there is an effective, single parameter MRI or fewer combinations of them, it would be worthwhile searching for them. In this study, we calculated the best overall accuracies and AUCs of all combinations of T1, T1c, T2, and FLAIR MRIs to see which combination resulted in the best classification. The results are shown in Table 2.

From Table 2, one can observe that the best classifications are made when the random forest utilizes all four MRIs. Additionally, it can be deduced that T1c MRIs contain some information that is more informative in the classification task because of the overall increase in accuracy and AUC in those combinations that contain T1c images. Interestingly, while the addition of more modalities increased the accuracy and AUC overall, T1c alone is as good as most combinations of MR images.

Specifically, when looking at the features extracted at each combination of MRIs a few interesting observations appear. First, our results show that in single MR imaging that takes advantage of the spin-lattice relaxation or longitudinal relaxation times, T1 and T1c, the features that were selected to be the most informative, outside of shape features, were intensity features. On the other hand, for MR imaging that takes advantage of the spin-spin relaxation or transverse relaxation times, T2 and FLAIR, Gray-Level Co-occurrence Matrix (GLCM) features were selected to be informative features in addition to the shape and intensity features. This suggests that: 1) GLCM features are important for those MRIs that take advantage of the T2 relaxation effects and 2) that intensity features are important to the correct classification of gliomas largely due to their selection in T1c features as discussed previously. This is supported in the introduction of another MRI modality in the classification task where the most significant differences in features between the two categories, T1 relaxation imaging vs. T2 relaxation imaging, are either intensity features or GLCM features respectively. A more detailed information about the features selected for the various combinations are shown in the Table 3.

The second observation we observed was that initially, shape features were the most important in the classification task as observed historically within the clinical perspective. However, while sufficient shape features may be required initially, to increase the accuracy of the classification from adequate to something exciting, additional features like intensity features or GLCM features are required for fine-tuning the accuracy. This was observed in our findings that as the features and MRIs increased, the number and ratio of shape features to other features decreased. A parallel could be drawn from how the disease is classified clinically and how the machine learning method classifies. Generally, easily observable and inherently informative shape features could allow physicians to draw a quick conclusion on the disease grade. However, additional information such as histopathology reports can better inform the party into drawing the most accurate conclusion. Similarly, the machine uses the most cost effective features, namely shape features, to quickly draw a fairly accurate conclusion while supporting features such as GLCM and intensity features allow the fine tuning of the conclusion.

**Table 3.** The adding features for all modality combination with a correlation threshold from 0.10 to 0.15.

| All MRIs | Feature Groups | Feature |
|---|---|---|
| Flair | GLCM(3D) | Correlation(4) InformationMeasureCorr1(1) InformationMeasureCorr2(1) |
| | Neighbor Intensity Difference(3D) | Coarseness(1) |
| | Shape | Spherical Disproportion(1) Surface Area(1) |
| T1 | Intensity Direct | Local Entropy Std(1) |
| | Shape | Spherical Disproportion(1) Surface Area(1) |
| T1c | Shape | Spherical Disproportion(1) Surface Area(1) |
| T2 | GLCM(3D) | InformationMeasureCorr2(1) |
| | Shape | Spherical Disproportion(1) Surface Area(1) |

### 4.2 Results across correlation thresholds

While comparing the prediction accuracies over correlation thresholds, as shown in Figure 3, there are two areas of interest. The first at thresholds from 0.1 to 0.15 and the second from 0.15 to 0.95. In the first area, there is a major increase in the prediction accuracy and AUC while in the second, the accuracy and AUC remained relatively stable. To understand these changes, we considered the features that were added or subtracted in those two areas. For the first, the features added from 0.1 to 0.15 are shown in Table 3. In the second area, there are more features with an average of more than fifty features. One correlation threshold with 0.575 has 52 features and are listed on Table 4.

In Table 4, we can see that shape features are in common among all four modalities. This is consistent with several other studies which have reported shape features to effectively characterize and classify glioma phenotype [28-30]. Another study demonstrated Shape-Compactness radiomic signature to have great prognostic performance decoding tumor phenotype on the lung cancer datasets and head-and-neck cancer data sets (12), which indicated shape features are more generalizable features. On the other hand, different MRIs have different features that were significant and non-redundant. Particularly, from our previous observation of T1c MRIs contributing more in improving overall accuracy and AUC, it seems that T1c intensity features specifically contribute more to the overall accuracy and AUC. While for other parameter MRIs like FLAIR, GLCM textural features contribute more. This is supported by papers as early as 1973 [31]and later confirmed for the use in radiomic applications by Ravanelli et al. [32] in 2010 for non-small cell lung cancer (NSCLC) and Aerts et al. (12) in 2014 for a more general application. While these papers used mono-modality radiomic features to discover radiomic signatures of lung and head and neck cancers, our study shows that the combination of these features from different parametric MRIs may be an even stronger radiomic signature, or finger print, for classifications of gliomas and could allow future studies to test with and against these features to develop a robust classification or predictive model. With these in mind, future studies could focus on finding specific features for specific parameter MRIs or modalities to improve their tasks. One limitation of our study is the lack of patient age within the BRATs dataset which has previously been reported to affect feature extraction between two groups of gliomas patients, <45 and ≥45 years old [33]. Clinically, this indicates that different age groups may have different origins and phenotypic expressions. With respect to the proposed classification method, this indicates that to more accurately classify patients, stratification of patients based on age groups are required to build the best radiomic feature patterns that differentiate HGG and LGG.

**Table 4.** The features at the correlation threshold of 0.575 for all modality combination.

| All MRIs | Features not in common | Features in common |
|---|---|---|
| Flair | GLCM(15)<br>Intensity (3)<br>Shape (1) | Shape-Compactness2<br>Shape-Spherical disproportion<br>Shape-Sphericity<br>Shape-Surface Area |
| T1 | Gradient Orient Histogram(1)<br>Intensity(3)<br>GLCM(3) | |
| T1c | Intensity(4) | |
| T2 | GLCM(4)<br>Intensity(2) | |

## 5. CONCLUSION

In conclusion, our method of using multi-parametric radiomic features with machine learning methods performed well in classifying HGG and LGG patients in the BRATs dataset with an average of 91% accuracy. This study shows that radiomic features of multi-parametric MRI, especially T1c intensity features, are highly correlated with tumor grades and may further elucidate the underlying tumor biology and response to therapy. The radiomic analysis combined with machine learning as demonstrated in this study can be generalized to a broader set of diseases and combinations of imaging modalities such as combining radiomic features of CT, PET, and MR images of a disease. This study showed that while certain MR images performed better, it was the combination of all MR images that gave the best results. This method of classification will be tested further as more the BRATs database gets updated. With advances in imaging, machine learning, and radiomic analysis, a more robust, accurate, and validated classification model will be available in the future to assist oncologists in delivering better and more personalized treatments.

**ACKNOWLEDGEMENTS**

This work was supported in parts by grants from National Institutes of Health [R01CA215718 to XY, R01CA203388 and R01CA169937 to HM] and Emory Winship Cancer Institute pilot grant.